%% file: main.tex
\documentclass{article}
\usepackage[utf8]{inputenc}

\usepackage[utf8]{inputenc} 
\usepackage[T1]{fontenc}    
\usepackage{hyperref}       
\usepackage{url}            
\usepackage{booktabs}       
\usepackage{amsfonts}       
\usepackage{nicefrac}       
\usepackage{microtype}      
\usepackage{graphicx}
\usepackage{subcaption}
\usepackage{amsmath}
\usepackage{algorithm}
\usepackage{algorithmic}
\usepackage{bm}
\usepackage{wrapfig}

\usepackage[
backend=biber,
style=numeric,
sorting=nyt 
]{biblatex}

\bibliography{main}

\usepackage{tikz-cd}
\usetikzlibrary{positioning}
\usetikzlibrary{shapes}
\usepackage{float}
\usepackage{tkz-euclide}
\usetkzobj{all}
\usepackage{enumitem}
\usepackage{array}
\usepackage{tabularx}

\newcommand\setrow[1]{\gdef\rowmac{#1}#1\ignorespaces}
\usepackage{caption}
\usepackage{subcaption}

\usepackage[super]{nth}

\newcommand{\tbsmlocation}{https://github.com/facebookresearch/tbsm}

\title{Time-based Sequence Model for Personalization and Recommendation Systems}

\author{
  Tigran Ishkhanov, Maxim Naumov, Xianjie Chen, \\ Yan Zhu, Yuan Zhong, Alisson Gusatti Azzolini, Chonglin Sun,  \\ Frank Jiang, Andrey Malevich and Liang Xiong \\
  Facebook, 1 Hacker Way, Menlo Park, CA 94065 \\
  \texttt{\{tishkhanov,mnaumov,cxj\}@fb.com}
}
\date{}

\begin{document}

\maketitle

\begin{abstract}
In this paper we develop a novel recommendation model that explicitly incorporates time information. The model relies on an embedding layer and TSL attention-like mechanism with inner products in different vector spaces, that can be thought of as a modification of multi-headed attention. This mechanism allows the model to efficiently treat sequences of user behavior of different length. We study the properties of our state-of-the-art model on statistically designed data set. Also, we show that it outperforms more complex models with longer sequence length on the Taobao User Behavior dataset. 
\end{abstract}

\input{introduction}

\input{model}

\input{data}

\input{experiments}

\input{related}

\input{conclusion}

\subsubsection*{Acknowledgments}
The authors would like to thank Mikhail Smelyanskiy for supporting this work.  

\printbibliography[]

\end{document}

%% file: introduction.tex
\section{Background}

Recommendation systems play an important role in many e-commerce applications as well as search and ranking services \cite{covington2016, he2014practical, lake2019largescale, naumov2020, park2018, qipi2019, wang2019sequential, zhu2019joint}. There are two main strategies to perform recommendations: content and collaborative filtering. In content filtering the user creates a profile based on its interest, while human experts create a profile for the product. An algorithm matches the two profiles and recommends the closest matches to the user. For example, this approach is taken by the Pandora Music Genome Project \cite{mgp}.

In collaborative filtering, the recommendations are based only on user past behavior from which the future behavior is derived. The advantage of this approach is that it requires no external information and is not domain specific. The challenge is that in the beginning very few user-item interactions are available. For instance, this cold start problem is addressed by Netflix by asking the user for a few favorite movies when creating their profile for the first time \cite{netflix}.

\begin{figure}[h!]
    \centering
    \begin{subfigure}[b]{.3\linewidth}
        \includegraphics[width=\linewidth]{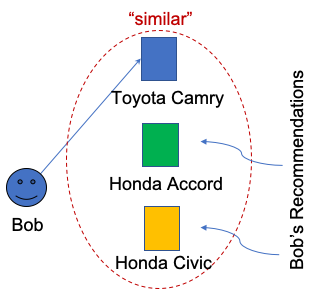}
        \caption{Item-oriented}
    \end{subfigure}
    \quad
    \begin{subfigure}[b]{.3\linewidth}
        \includegraphics[width=\linewidth]{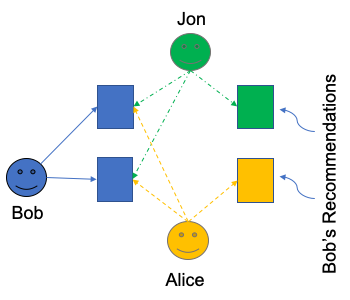}
        \caption{User-oriented}
    \end{subfigure}
    \quad
    \begin{subfigure}[b]{.3\linewidth}
        \includegraphics[width=\linewidth]{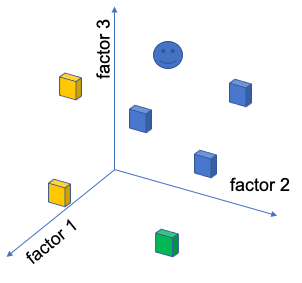}
        \caption{Implicit factor ($d=3$)}
    \end{subfigure}
    \caption{Neighborhood and implicit factor collaborative filtering schemes.}
    \label{fig:user_item_neighborhoods}
\end{figure}

Further, the collaborative filtering schemes are often split into neighborhood and latent factor methods. The neighborhood methods either find clusters of items (item-oriented) or identify groups of users that share some items (users-oriented) and then recommend new items to a particular user based on it \cite{ning2015}. In contrast, latent factor methods characterize users and items based on implicit factors to be determined by the algorithm/model. The $d$ factors can often be thought of as elements of a vector in $\mathbb{R}^{d}$ \cite{datta2017, ilic2015, rossetti2013}. We show a simple schematic drawing of these methods on Fig. \ref{fig:user_item_neighborhoods}, where circles and rectangles denote users and items, respectively.   

The most common example of implicit factor methods is matrix factorization \cite{koren2009matrix}. It attempts to solve the matrix completion problem that in its simplest form can be formulated as following. Let $r_{ui}$ be the rating of $i$-th item by $u$-th user. We would like to find an approximation to partially filled-in matrix of user-item interactions $R = [r_{ui}] \in \mathbb{R}^{m \times n}$ as
\begin{equation}
R \approx WV^T
\end{equation}
where $W^T=[\textbf{w}_{1},...,\textbf{w}_{m}]$ and $V^T=[\textbf{v}_{1},...,\textbf{v}_{n}]$ can be though of as embedding tables. Note that individual vectors $\textbf{w}_{u} \in \mathbb{R}^d$ and $\textbf{v}_{i} \in \mathbb{R}^d$ can be interpreted as representations of $u$-th user and $i$-th item, respectively. 

This problem can be solved by finding 
\begin{equation}
\min_{W,V} \sum_{(u,i) \in \mathcal{S}} (r_{ui} - \textbf{w}_u^T\textbf{v}_{i})
\label{eq:matrixfactorization}
\end{equation}
where $\mathcal{S}$ is a set of known ratings. 

Notice that this formulation can readily be generalized by exchanging the dot product $\textbf{w}_u^T\textbf{v}_{i}$ in \eqref{eq:matrixfactorization} for a more complex interaction function $\phi(u,i;\theta)$, with input indices $u$, $i$ and parameters $\theta$ \cite{he2017neural}. Moreover, we can often provide a richer context with continuous values $\textbf{x}$ as well as discrete sets of indices $\{u_1,...,u_p\}$ for the user and $\{i_1,...,i_q\}$ for the item features, resulting in 

\begin{equation}
\min_{\theta} \sum_{(u,i) \in \mathcal{S}} (r_{ui} - \phi(\textbf{x}, u_1,...,u_p,i_1,...,i_q;\theta))
\label{eq:dlrm}
\end{equation}
Typically these indices are used for embedding table lookups \cite{naumov2019}, with results propagated further. In effect many of the state-of-the-art recommendation models that treat time implicitly adopt this formulation and focus on the design of the function $h$ that defines a particular model \cite{ cheng2016wide, guo2017deepfm, kuchaiev2017training,  lian2018xdeepfm, naumov2019deep, sedhain2015autorec, wang2017deep}. 

The recommendation models can also explicitly incorporate timing component. In this scenario we often have a set of events $\mathcal{E} = \{u,i,r_{ui},t\}$ for time $t=1,..,\tau-1$ and we are interested in predicting the next event at time $\tau$.  Therefore, the dataset to train such models is often organized into time series corresponding to history of user activity, rather than simply a collection of user and item features.

The models that implicitly incorporate timing component can often produce a representation of an event as an embedding vector, therefore making it possible to generate a sequence of such embeddings $Z = [\textbf{z}_1, ..., \textbf{z}_{\tau-1}]$. This sequence can be generated independently one item at a time, using recurrent neural networks (RNNs) or some other mechanism represented by function $\psi$ in

\begin{equation}
\textbf{z}_t = \psi(\textbf{x}, u_1,...,u_p, i_1^{(t)},...,i_q^{(t)},\textbf{z}_{t-1};\theta) 
\label{eq:sequence_emb}
\end{equation}
for $t=1,...,\tau-1$ and $\textbf{z}_0$ being some initial state. Then, the prediction of the new rating relies on its context as well as its relevance relative to the earlier events, which may be represented through an attention mechanism \cite{NIPS2017_7181}, in  
\begin{eqnarray}
\textbf{z}_{\tau} &=& \psi(\textbf{x}, u_1,...,u_p,i_1^{(\tau)},...,i_q^{(\tau)},\textbf{z}_{\tau-1};\theta) \label{eq:tbsm1} \\
\textbf{c} &=&  \texttt{attn}(\textbf{z}_{\tau}, Z) \label{eq:tbsm2} \\
\min_{\theta} &\sum_{(u,i) \in \mathcal{S}}& (r_{ui} - \varphi(\textbf{z}_{\tau}, \textbf{c};\theta))
\label{eq:tbsm3}
\end{eqnarray}
The implementation details specifying interaction of function $\psi$ that constructs embedding of prior events, attention mechanism \texttt{attn} and function $\varphi$ that computes the final rating define individual models\footnote{Note that for a sequence of $\tau=1$, and no context $\textbf{c}$ from attention mechanism, the model \eqref{eq:tbsm1} - \eqref{eq:tbsm3} reduces to the one without an explicit timing component in \eqref{eq:dlrm}, so that $\phi = \varphi \circ \psi$.} \cite{qipi2019, zhou2018deepi, zhou2018deep}.

%% file: model.tex
\section{Time-based Sequence Model}\label{model}

In this section we introduce the time-based sequence model (TBSM) for user behavior data as well as its different variations which serve to demonstrate several plausible approaches to this problem. In subsequent discussion we define a datapoint as a sequence of events corresponding to a particular user. 

\subsection{Embedding Layer}

As a first step we pass each datapoint representing a sequence of events for a given user through an embedding layer that generates an embedding vector for each event. The use of an embedding layer is a standard step in deep learning solutions to the CTR problems \cite{chen2019flen, guo2017deepfm, zhou2018deepi}.

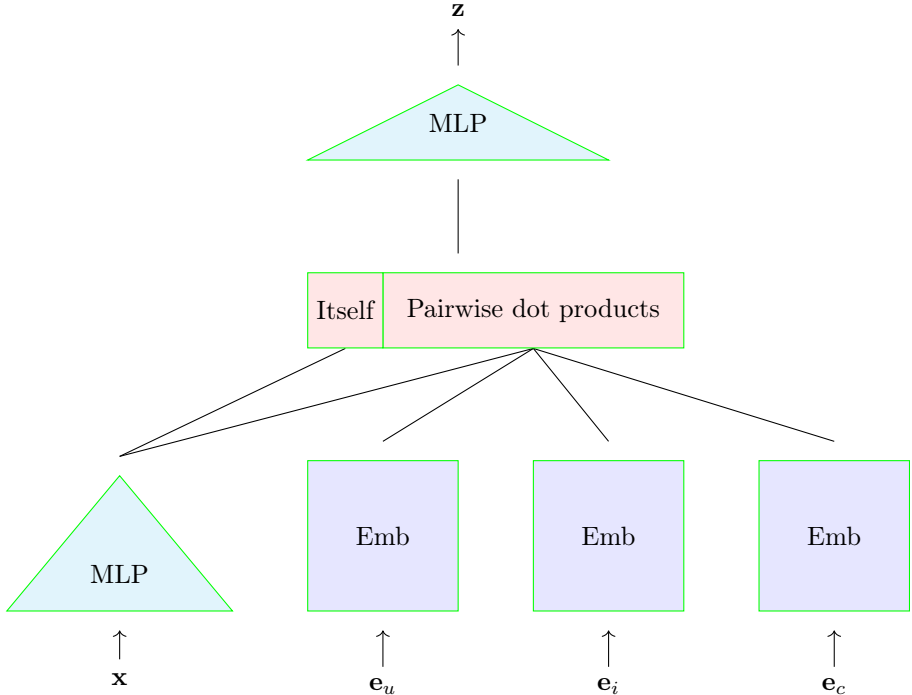
\begin{figure}[!tbp]
\centering
\begin{tikzpicture}[
emptynode/.style={minimum size=5mm},
]

\node[emptynode]    (f0)   at  (1.5,-0.90) {$\textbf{x}$};
\node[emptynode]    (f1)   at  (5,-1) {$\textbf{e}_{u}$};
\node[emptynode]    (f2)   at  (8,-1) {$\textbf{e}_{i}$};
\node[emptynode]    (f3)   at  (11,-1) {$\textbf{e}_{c}$};

\node[emptynode]    (f01)   at  (1.5,0) {};
\node[emptynode]    (f11)   at  (5,0) {};
\node[emptynode]    (f21)   at  (8,0) {};
\node[emptynode]    (f31)   at  (11,0) {};

\draw[->] (f0.north) -- (f01.south);
\draw[->] (f1.north) -- (f11.south);
\draw[->] (f2.north) -- (f21.south);
\draw[->] (f3.north) -- (f31.south);

\filldraw[fill=cyan!10!white, draw=green] (0,0) node[anchor=north]{}
  -- (3,0) node[anchor=north]{}
  -- (1.5,1.8) node[anchor=south]{}
  -- cycle;
\node[emptynode]    (t0)   at  (1.5,0.5) {MLP};

\filldraw[fill=blue!10!white, draw=green] (4,0) rectangle (6,2);
\filldraw[fill=blue!10!white, draw=green] (7,0) rectangle (9,2);
\filldraw[fill=blue!10!white, draw=green] (10,0) rectangle (12,2);
\node[emptynode]    (t1)   at  (5,1) {Emb};
\node[emptynode]    (t2)   at  (8,1) {Emb};
\node[emptynode]    (t2)   at  (11,1) {Emb};

\node[emptynode]    (d1)   at  (1.5,1.8) {};
\node[emptynode]    (s1)   at  (5,2) {};
\node[emptynode]    (s2)   at  (8,2) {};
\node[emptynode]    (s3)   at  (11,2) {};
\node[emptynode]    (mg)   at  (7,3.75) {};
\node[emptynode]    (mg_itself)   at  (4.5,3.75) {};
\draw[-] (d1.north) -- (mg_itself.south);
\draw[-] (d1.north) -- (mg.south);
\draw[-] (s1.north) -- (mg.south);
\draw[-] (s2.north) -- (mg.south);
\draw[-] (s3.north) -- (mg.south);

\filldraw[fill=red!10!white, draw=green] (5,3.5) rectangle (9,4.5);
\filldraw[fill=red!10!white, draw=green] (4,3.5) rectangle (5,4.5);
\node[emptynode]    (t3)   at  (7,4) {Pairwise dot products};
\node[emptynode]    (t3)   at  (4.5,4) {Itself};
\node[emptynode]    (p)   at  (6,4.5) {};

\filldraw[fill=cyan!10!white, draw=green] (4,6) node[anchor=north]{}
  -- (8,6) node[anchor=north]{}
  -- (6,7) node[anchor=south]{}
  -- cycle;
\node[emptynode]    (t4)   at  (6,6.5) {MLP};

\node[emptynode]    (m0)   at  (6,6) {};
\draw[-] (p.north) -- (m0.south);
\node[emptynode]    (m1)   at  (6,7) {};

\node[emptynode]    (m2)   at  (6,7) {};
\node[emptynode]    (final)   at  (6,8) {$\textbf{z}$};
\draw[->] (m2.north) -- (final.south);

\end{tikzpicture}
\caption{DLRM embedding layer architecture} 
\label{fig:DLRM}
\end{figure}

We choose to use deep learning recommendation model (DLRM) \cite{naumov2019deep} to express this embedding layer\footnote{Note that we stop DLRM one layer prior to the final output, where we have an embedding representation and before it is converted to a probability of a click by last layer of the model.}. It accepts input continuous features \textbf{x}, a set of discrete features $\textbf{e}_s$ and generates an output embedding $\textbf{z}$. In particular, in our experiments we will be interested in three one-hot categorical features corresponding to user $\textbf{e}_u$, item $\textbf{e}_i$ and category $\textbf{e}_c$ as well as a single continuous feature $\textbf{x}$ corresponding to time, as shown on Fig. \ref{fig:DLRM}. Notice that in DLRM categorical features get embedded into $\mathbb{R}^d$ using $\textbf{w}_s^T = \textbf{e}_{s}^TW$, 
where $W$ is a trainable weight parameter. On the other hand, continuous features are passed through the bottom multilayer perceptron (MLP), whose final dimension is the same as embedding dimension of all categorical features. In the following step we take all pairwise dot products between dense and categorical features. The final step of DLRM takes these dot products combined with embedded dense feature and passes it through the top MLP resulting in a vector $\textbf{z} \in \mathbb{R}^n$. 

As a result of the application of DLRM to each position in the input sequence our output can be thought of as a time series in $\mathbb{R}^n$. We use notation $\textbf{z}_t$ for DLRM output of item at position $t$, and $\textbf{z}_{\tau}$ for the output of the last item.

\subsection{Time Series Layer}
\begin{figure}[!tbp]
\centering
\begin{tikzpicture}[
emptynode/.style={minimum size=5mm},
]

\node[emptynode]    (i0)   at  (0,0) {$\textbf{z}_0$};
\node[emptynode]    (i1)   at  (2,0) {$...$};
\node[emptynode]    (i2)   at  (4,0) {$\textbf{z}_{\tau-1}$};
\node[emptynode]    (j)   at  (8,0) {$\textbf{z}_{\tau}$};

\node[emptynode]    (D1)   at  (4,2.1) {};

\draw[-] (i0.north) -- (D1);
\draw[-] (i1.north) -- (D1);
\draw[-] (i2.north) -- (D1);
\draw[-] (j.north) -- (D1);

\filldraw[fill=red!10!white, draw=green] (2,2) rectangle (6,3);
\node[emptynode]    (t3)   at  (4,2.5) {Similarities $\langle\textbf{z}_t,\textbf{z}_j\rangle$};
\node[emptynode]    (rt)   at  (4,2.75) {};
  
\filldraw[fill=cyan!10!white, draw=green] (2,4) node[anchor=north]{}
  -- (6,4) node[anchor=north]{}
  -- (4,5) node[anchor=south]{}
  -- cycle;
\node[emptynode]    (t0)   at  (4,4.4) {MLP};
\node[emptynode]    (mb)   at  (4,4.25) {};
\node[emptynode]    (mn)   at  (4,5) {};
\node[emptynode]    (a)   at  (4,6) {$\textbf{c}=Z\textbf{a}$};

\draw[-] (rt) -- (mb.south);
\draw[->] (mn.north) -- (a.south);

\end{tikzpicture}
\caption{TSL architecture} 
\label{fig:TS}
\end{figure}
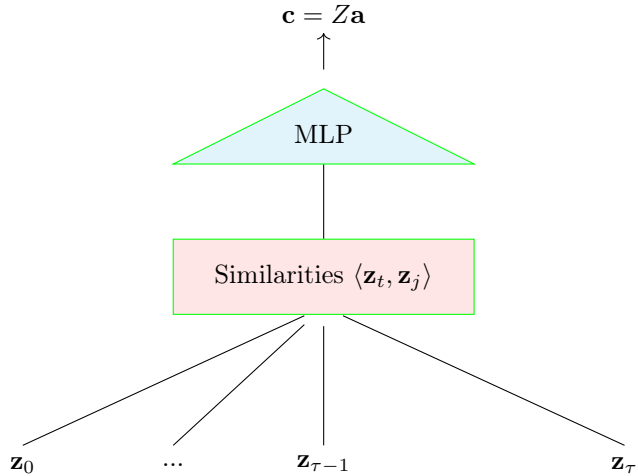

The time series layer (TSL) is designed specifically for time series processing. The high-level purpose of this layer is to relate the time series vectors to the last item. In this layer we compute the similarities between the sequence of vectors $Z=[\textbf{z}_1,\ldots,\textbf{z}_{\tau-1}]$ and the last item $\textbf{z}_{\tau}$, pass them though a MLP to create coefficients $\textbf{a}$ and produce a context vector $\textbf{c}=Z\textbf{a}$ as shown in Fig. \ref{fig:TS}. We will use several such layers when composing the full model in the next section. Also, we point out that this scheme resembles an attention mechanism, with a few small but important modifications that will be discussed next.  

The first modification is that the vectors $\textbf{z}_t$ are normalized by projecting them onto a unit sphere $\mathbb{S}^n$:
\begin{equation}
\textbf{z}_t = \textbf{z}_t / ||\textbf{z}_t||_2
\label{eq:proj_unit_sphere}
\end{equation}
Also, we replace the dot product by one or more positive definite inner products \begin{equation}
\langle \textbf{z}_t, \textbf{z}_{\tau} \rangle  = (A\textbf{z}_t)^T (A\textbf{z}_{\tau}) = \textbf{z}_t^T (A^TA) \textbf{z}
\label{eq:inner_prod}
\end{equation}
for some nonsingular matrix $A \in \mathbb{R}^{n \times n}$. The inner product achieves two inter-dependent goals: (i) capturing label dependence on coupling between time series in different coordinates of the embedded space, and (ii)  learning different similarity measures in the embedded space. 

Note that in combination with spherical projection dot product becomes a cosine similarity, a measure widely used in natural language processing (NLP) \cite{sitikhu2019comparison}. We also may view inner product between projected vectors as a more general similarity measure as compared to cosine similarity. We remark that the connection with NLP can be elucidated if we think of analogy of item vocabulary to word vocabulary, and probability of last item corresponding to probability of next word under a specific language model. The difference with NLP case is the size of vocabulary, which in our case can easily exceed tens of millions of items. 

Let us now point out how our attention-like mechanism is different from multi-head attention \cite{NIPS2017_7181}. Recall that in multi-head attention  the similarity between key and query is constructed via 
\begin{equation}
\langle \textbf{w},\textbf{v} \rangle = (F \textbf{w})^T (G \textbf{v}),
\end{equation}
where $F$ and $G$ are matrices whose dimensions are chosen so that their outputs have same dimensionality. We remark that in our case we do not need to map a pair of vectors into a common space since they are already in the output space of DLRM embedding layer. Therefore, there are three major differences of our approach as compared to multi-head attention. First, there are fewer trainable parameters since it involves only one matrix $A$. Second, the weighting has a clear geometric interpretation since cosine similarity is directly related to a distance function on the unit sphere
\begin{equation}
    D(\textbf{w},\textbf{v}) = \frac{\cos^{-1}(\langle \textbf{w}, \textbf{v} \rangle )}{\pi}
\end{equation}
Finally, when using multiple TSLs, we will pair them with individual MLPs, while in multi-headed attention a single shared MLP is used for all. While these differences seem small, we will show in our experiments that they result in major improvement in the statistical performance of the model.

\subsection{TBSM: Composing All Layers Together}

Let us now compose all layers together into a single model, that will be referred to as time-based sequence model (TBSM).

Notice that the output of embedding layer corresponding to \eqref{eq:tbsm1} and TSL corresponding to \eqref{eq:tbsm2} can be concatenated and supplied as an input $[\textbf{z}_{\tau},\textbf{c}]$ to the MLP which would produce probability of a click $p$. 

Further, we may choose to use multiple TSLs, with different similarity measures between vectors, or even using different sequence lengths for each of them. In this case, we obtain $k$ concatenated outputs $[\textbf{z}_{\tau},\textbf{c}_i]$ and pass each of them through an independent MLP, resulting in a distinct probability $p_i$ for $i=1,...,k$. Once again note that this approach resembles the use of multiple heads in an attention mechanism \cite{transformer, shazeer2019fast, voita2019analyzing,  attention_attention}, but aside from the use of spherical projection in \eqref{eq:proj_unit_sphere} and well defined inner products in \eqref{eq:inner_prod}, the output uses individual rather than shared MLPs. 

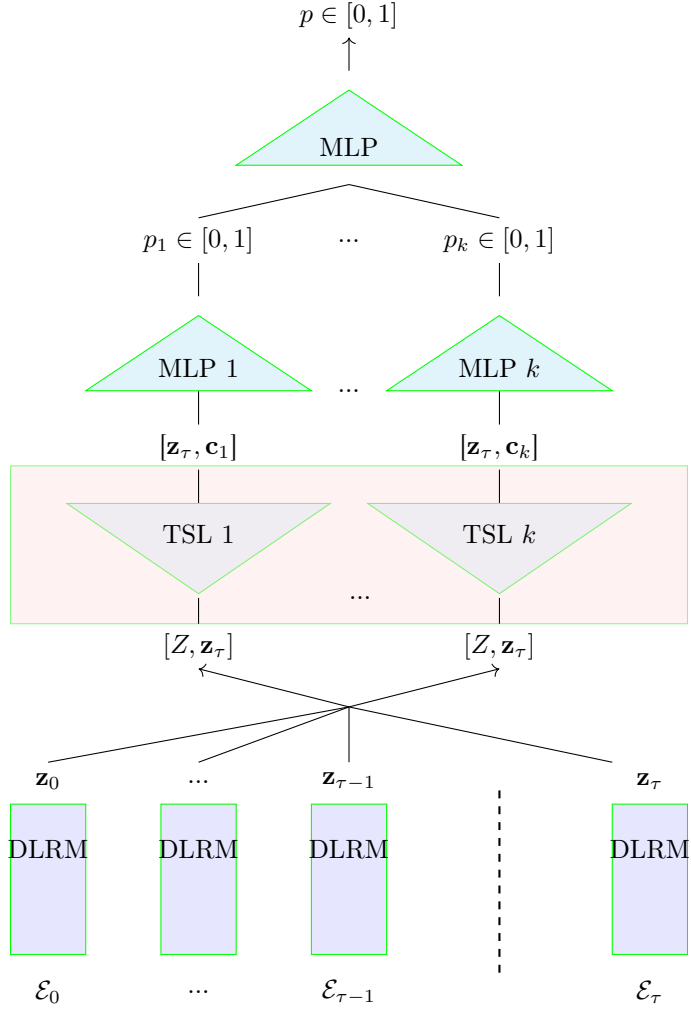
\begin{figure}[!ht]
\centering
\begin{tikzpicture}[
roundnode/.style={circle, draw=red!30, fill=red!5, very thick, minimum size=5mm},
roundnode2/.style={circle, draw=green!30, fill=cyan!5, very thick, minimum size=5mm},
emptynode/.style={minimum size=5mm},
]

\filldraw[fill=blue!10!white, draw=green] (-1,-1) rectangle (0,1);
\filldraw[fill=blue!10!white, draw=green] (1,-1) rectangle (2,1);
\filldraw[fill=blue!10!white, draw=green] (3,-1) rectangle (4,1);
\filldraw[fill=blue!10!white, draw=green] (7,-1) rectangle (8,1);

\node[emptynode]    (z0)   at  (-0.5,-1.5) {$\mathcal{E}_0$};
\node[emptynode]    (z1)   at  (1.5,-1.5) {$...$};
\node[emptynode]    (z2)   at  (3.5,-1.5) {$\mathcal{E}_{\tau-1}$};
\node[emptynode]    (z3)   at  (7.5,-1.5) {$\mathcal{E}_{\tau}$};

\node[emptynode]    (z0)   at  (-0.5,0.40) {DLRM};
\node[emptynode]    (z0)   at  (1.5,0.40) {DLRM};
\node[emptynode]    (z0)   at  (3.5,0.40) {DLRM};
\node[emptynode]    (z0)   at  (7.5,0.40) {DLRM};

\node[emptynode]    (r0)   at  (-0.5,1.3) {$\textbf{z}_{0}$};
\node[emptynode]    (r1)   at  (1.5,1.3) {$...$};
\node[emptynode]    (r2)   at  (3.5,1.3) {$\textbf{z}_{\tau-1}$};
\node[emptynode]    (r3)   at  (7.5,1.3) {$\textbf{z}_{\tau}$};
\node[emptynode]    (r31)   at  (7,1.3) {};

\node[emptynode]    (mg)   at  (3.5,2.55) {};
\node[emptynode]    (ts1)   at  (1.5,3.1) {$[Z,\textbf{z}_{\tau}]$};
\node[emptynode]    (ts2)   at  (5.5,3.1) {$[Z,\textbf{z}_{\tau}]$};

\draw[-] (r0.north) -- (mg.south);
\draw[-] (r1.north) -- (mg.south);
\draw[-] (r2.north) -- (mg.south);
\draw[-] (r31.north) -- (mg.south);

\draw[->] (mg.south) -- (ts1.south);
\draw[->] (mg.south) -- (ts2.south);

\filldraw[fill=cyan!10!white, draw=green] (3.75,5) node[anchor=north]{}
  -- (7.25,5) node[anchor=north]{}
  -- (5.5,3.8) node[anchor=south]{}
  -- cycle;
  
\filldraw[fill=cyan!10!white, draw=green] (-0.25,5) node[anchor=north]{}
  -- (3.25,5) node[anchor=north]{}
  -- (1.5,3.8) node[anchor=south]{}
  -- cycle;
  
\filldraw[fill=red!10!white, draw=green, opacity=0.5] (-1,3.4) rectangle (8,5.5);
\node[emptynode]    (mt2)   at  (3.65,3.75) {...};
  
\node[emptynode]    (mt1)   at  (1.5,4.6) {TSL $1$};
\node[emptynode]    (mt2)   at  (5.5,4.6) {TSL $k$};

\node[emptynode]    (ce1)   at  (1.5,5.75) {[$\textbf{z}_{\tau},\textbf{c}_1$]};
\node[emptynode]    (ce2)   at  (5.5,5.75) {[$\textbf{z}_{\tau},\textbf{c}_k$]};

\node[emptynode]    (mt1n)   at  (1.5,4.75) {};
\node[emptynode]    (mt2n)   at  (5.5,4.75) {};

\draw[-] (mt1n.north) -- (ce1.south);
\draw[-] (mt2n.north) -- (ce2.south);

\node[emptynode]    (mtt1)   at  (1.5,4) {};
\node[emptynode]    (mtt2)   at  (5.5,4) {};
\draw[-] (ts1.north) -- (mtt1.south);
\draw[-] (ts2.north) -- (mtt2.south);

\filldraw[fill=cyan!10!white, draw=green] (0,6.5) node[anchor=north]{}
  -- (3,6.5) node[anchor=north]{}
  -- (1.5,7.5) node[anchor=south]{}
  -- cycle;
  
\filldraw[fill=cyan!10!white, draw=green] (4,6.5) node[anchor=north]{}
  -- (7,6.5) node[anchor=north]{}
  -- (5.5,7.5) node[anchor=south]{}
  -- cycle;
  
\node[emptynode]    (mlp1s)   at  (1.5,6.75) {};
\node[emptynode]    (mlp2s)   at  (5.5,6.75) {};

\node[emptynode]    (mlp1n)   at  (1.5,7.5) {};
\node[emptynode]    (mlpdots)   at  (3.5,6.5) {...};
\node[emptynode]    (mlp2n)   at  (5.5,7.5) {};

\node[emptynode]    (z1)   at  (1.5,8.5) {$p_1\in [0,1]$};
\node[emptynode]    (zdots)   at  (3.5,8.5) {...};
\node[emptynode]    (z2)   at  (5.5,8.5) {$p_k\in [0,1]$};

\draw[-] (mlp1n.north) -- (z1.south);
\draw[-] (mlp2n.north) -- (z2.south);

\filldraw[fill=cyan!10!white, draw=green] (2,9.5) node[anchor=north]{}
  -- (5,9.5) node[anchor=north]{}
  -- (3.5,10.5) node[anchor=south]{}
  -- cycle;
  
\node[emptynode]    (fmlps)   at  (3.5,9.5) {};

\draw[-] (z1.north) -- (fmlps.south);
\draw[-] (z2.north) -- (fmlps.south);
 
\node[emptynode]    (p1)   at  (1.5,6.8) {MLP $1$};
\node[emptynode]    (p2)   at  (5.5,6.8) {MLP $k$};

\node[emptynode]    (t)   at  (3.5,9.75) {MLP};

\draw[-] (ce1.north) -- (mlp1s.south);
\draw[-] (ce2.north) -- (mlp2s.south);

\node[emptynode]    (fmlpn)   at  (3.5,10.5) {};

\node[emptynode]    (p)   at  (3.5,11.5) {$p\in [0,1]$};

\draw[->] (fmlpn.north) -- (p.south);

\node[emptynode]    (sep1)   at  (5.5,-1.5) {};
\node[emptynode]    (sep2)   at  (5.5,1.5) {};

\draw[thick,dashed] (sep1.north) -- (sep2.south);

\end{tikzpicture}
\caption{TBSM architecture with $k$ TSLs} 
\label{fig:fullmodel}
\end{figure}

Finally, note that we can interpret each $p_i$ as a probability obtained by using the corresponding similarity measure. We can also think of them as being produced by an ensemble of models \cite{badirli2020gradient, Freund96experimentswith, tao2019deep}.  In this scenario, the last MLP determines their contribution to the final probability of click $p$. The comprehensive design of TBSM showing all of these components together is outlined on Fig. \ref{fig:fullmodel}.

%% file: data.tex
\section{Datasets}\label{datasets}

In this section we describe in detail the two datasets used in the experiments presented in this paper. We focus our attention on the Taobao User Behavior dataset\footnote{Taobao data can be found at https://tianchi.aliyun.com/dataset/dataDetail?dataId=649\&userId=1} \cite{zhu2019joint}, while we also design and use a custom synthetic dataset to explain the reasoning behind some of the architectural choices made in the model.

\subsection{Synthetic Dataset}\label{synthetic}

In synthetic dataset we let a datapoint be a time series, with each position in the series being a randomly generated vector in $\mathbb{R}^n$. This has a similar structure to Taobao train dataset after it passes through model's embedding layer, except that the binary label is made to explicitly depend on series data as follows.

Let us denote the first $\tau-1$ vectors in time series by $\textbf{z}_i$ and the last vector by $\textbf{z}_{\tau}$. Also, let $\tilde{\textbf{z}}$ denote a sum of $\delta$ (zero or more) vectors, each obtained by a random permutation of $n$ coordinates of $\textbf{z}_{\tau} \in \mathbb{R}^n$. Then, define the corresponding click/non-click label $l$ and function $f$ to be
\begin{eqnarray}
l &=& \texttt{sign}(f) \\  
f &=& \sum_{i=1}^{\tau-1}{f_1(\langle \textbf{z}_i,\textbf{z}_{\tau}\rangle) + f_2(\langle \textbf{z}_i,\tilde{\textbf{z}} \rangle)}
\label{eq:synthetic_function}
\end{eqnarray}
where $\langle .,. \rangle$ denotes an inner product. Note that the first term $f_1$ is a function of inner products between time series elements $\textbf{z}_i$ and $\textbf{z}_{\tau}$, while the second term $f_2$ contributes ``mixed'' products of coordinates in $\mathbb{R}^n$ (e.g. $\textbf{z}_{i}(1)\textbf{z}_{\tau}(2)$). We can control ``complexity'' of the dataset because we have a summand in $f_2$ for each permutation of vector $\textbf{z}_{\tau}$. If $f=f_1$, i.e. there are no mixed terms, then our time series in $\mathbb{R}^n$ can be thought of as $n$ independent time series in $\mathbb{R}^1$ that are completely decoupled from each other. The more summands we add into $f_2$ the stronger is coupling between time series in different coordinates. We will use this interplay to test different components of the model.

\subsection{Taobao User Behavior Dataset}

The raw Taobao User Behavior dataset has a total of about 4M items, 10K categories and 1M users. It is organized into a set of time series parameterized by a user. Each entry in the time series for a given user is a triple $(i,c,t)$, where user interacted with the item ``$i$'' belonging to category ``$c$'' at time ``$t$''\footnote{For example, ``soccer ball'' and ``sports'' could be a valid item-category pair.}. 

The processed Taobao User Behavior dataset is obtained from the raw dataset and organized into a set of datapoints, split across \texttt{taobao\_train.txt} and \texttt{taobao\_test.txt} files. We note that train dataset contains about 9M points, while test dataset contains a bit more than 296K points. A datapoint in processed dataset corresponds to a specific user and consists of 

\begin{enumerate}[label=(\roman*)]
\item user id $u$
\item pair of item and category id $(i, c)$ 
\item randomly generated (50/50 chance) binary label $l$ 
\item 200 pairs $(i, c)$ which encode user’s true behavior
\item 200 pairs $(i, c)$ randomly taken from the full dataset of 4M items, but that are different from user’s true behavior
\end{enumerate}

Notice that for a given user the sequence (iv) represent positive samples, i.e. clicks, while sequence (v) can be used to generate negative samples, i.e. non-clicks. Further, for users who have interacted with less than 200 items, the true behavior sequence is padded with zeros in front. In case user has longer than 200 item history, it is truncated at 200. The $200$-th point of true behavior sequence is special: if label $l=1$ the pair $(i, c)$ in (ii) is taken for true user history, otherwise if label $l=0$ the true point is replaced with randomly generated fake one taken from sequence (v). 

For instance, a sample datapoint (one line in text file) is shown below 
{\small
\begin{verbatim}
7 123 50 1   0,45,12,...123  0,17,89,...50   98,112,75,... 43,765,14,...     
\end{verbatim}
}
\noindent In this line user id $u=7$, pair of item and category id of $200$-th item is (123, 50), while label $l=1$. The sequence starting with 0,45 is 200 item ids from true history. The sequence starting with 0,17 is 200 are the category ids corresponding to the item ids in the first sequence. Note that this user only had 199 items in the true history sequence, therefore the first entry is padded by 0. Finally, last two sequences are item and category ids randomly generated for this user. 

For a fixed user behavior sequence length $\tau \le 200$, the train and test datasets are constructed using the following scheme. For the train dataset, positive datapoints (with label $l=1$) are constructed by taking contiguous subsequences of length $\tau$ from $200$-item long full sequence starting at random position. The negative datapoints (with label $l=0$) are constructed by replacing the item id following the chosen subsequence of length $\tau-1$ by an item chosen from the randomly generated 200-item sequence for this user. For the test dataset, we always take last $\tau$ items for each user with the provided label, so no replacements are made in this case. 

Finally we append user $u$ and time $t$ by taking $\tau$ equally-spaced values between $0$ and $1$ to each pair $(i, c)$ in length $\tau$ subsequence. Therefore, resulting in a final datapoint consisting of $\tau$ tuples $(u,i,c,l,t)$. Note that we think of each datapoint as a known time series of length $\tau-1$ and the last item whose probability we are trying to predict.

%% file: experiments.tex
\section{Experiments}\label{experiments}

\addtolength{\textheight}{+1cm} 

We perform all of our experiments on the following architecture of TBSM\footnote{TBSM code can be found at \tbsmlocation}. In embedding layer implemented through DLRM we let the embedding dimension $d=16$, bottom MLP to have a single layer $m \times d$ where $m$ is the number of dense features and top MLP have a single layer $\mu \times n$, where $\mu = d + (s)(s+1)/2$ is the number of (pair-wise) interactions between features and $s$ is the number of sparse features. Notice that for processed Taobao User Behavior dataset we have a single continuous feature time and we have three discrete features corresponding to user, item and category, therefore $m=1$ and $s=3$, respectively. We let $n=15$ and remark that reasonable choices for MLP layer sizes (e.g. $15$) do not have a measurable effect on performance and other values provide similar results. 

Inside TSL we let MLP have three layers, with dimensions $\tau \times n$, $n \times n$ and $n \times \tau$, where $\tau$ is the length of the time series. Note that input and output dimensions of this layer must coincide with $\tau$ because there are as many input values as there are events $\mathcal{E}_t$, and the output values provide coefficients in the linear combination $\textbf{c} = Z \textbf{a}$, where $\textbf{a}$ is the vector resulting from the MLP. In our experiments, we will show that in our model it will be sufficient to choose $\tau=20$ to achieve higher statistical performance than existing more complex models with larger $\tau$ that is supposed to be beneficial to them. 

Further, the MLP above TSL has dimensions $2n \times 4n$ and $4n \times 1$ since its input is a pair of $n$-dimensional vectors [$\textbf{z}_{\tau},\textbf{c}$]. When multiple TSLs are used, each with corresponding MLPs, these dimensions are replicated across them. 

Since our label is binary, it is natural to choose binary cross entropy
\begin{equation}
    \mathcal{L} = \sum y \log(p) + (1 - y) \log (1 - p)
\end{equation}
as model's loss function, where $y$ is the target and $p$ is the predicted probability. We also track accuracy as well as AUC metrics \cite{fawcett2006} throughout training. The model is trained for one epoch using Adagrad optimizer\footnote{We have experimented with SGD but the results were considerably worse with it when compared to Adagrad and therefore we do not report them here.} \cite{duchi2011}. We always report the average values over 10 runs to avoid different random initialization effects. 

\subsection{Synthetic Dataset}

In synthetic dataset the binary label encodes varying degree of coupling between coordinates in a time series vector. Let $\delta$ be the number of summands in $\tilde{\textbf{z}}$ in the second term in \eqref{eq:synthetic_function} which reflects it. Let us consider the model where TSL uses similarity $\langle .,. \rangle$ measured by standard single dot product or multiple inner products defined in \eqref{eq:proj_unit_sphere} and \eqref{eq:inner_prod}, as summarized in Tab. \ref{tab:per_synthetic}. 

\begin{table}[!h]
\caption{Performance (AUC) obtained on synthetic dataset}
\centering
\begin{tabular}{l c c c c}
\hline
\hline
Model & $\delta=0$ & $\delta=3$ & $\delta=12$ & $\delta=30$ \\
\hline
1-dot & 0.99 & 0.68 & 0.63 & 0.66 \\
1-inner & 0.82 & 0.60 & 0.58 & 0.67 \\
4-inner & 0.94 & 0.77 & 0.70 & 0.78 \\
8-inner & 0.98 & 0.80 & 0.79 & 0.80 \\
\hline
\end{tabular}
\label{tab:per_synthetic}
\end{table}

\begin{figure}[!htbp]
\centering
\includegraphics[width=7.4cm]{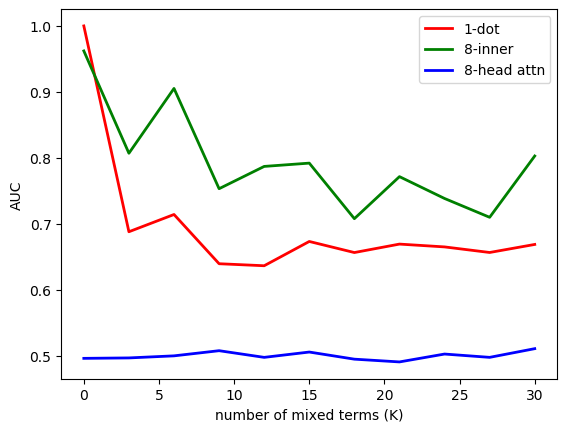}
\caption{Performance (AUC) of multi-head attention vs. dot and inner products} 
\label{fig:dot_inner}
\end{figure}

Notice that when there is no coupling between time series in different coordinates ($\delta=0$) the model using dot product in TSL has the best performance among the tested models. However, when strength of the coupling between coordinates increases (higher $\delta$) the performance for all models naturally deteriorates, but models with multiple inner products start to outperform the one with a single dot product, as shown on Fig. \ref{fig:dot_inner}. Notice also that in this experiment 8-head attention model is doing no better than a random model, while TSL significantly outperforms it.

\subsection{Taobao User Behavior Dataset}

Let us now study the behavior of different variations of TBSM. We consider different similarity measures, number and lengths of sequences as well as different attention and LSTM-based mechanisms for processing time series data.  

First, we experiment with different similarity measures in TSL. We let DotSim($\mathbb{R}^n$) refer to dot product in Cartesian coordinates, while GenSim($\mathbb{S}^n$) refer to inner product on a unit sphere computed using \eqref{eq:proj_unit_sphere} and \eqref{eq:inner_prod}. Also, we test the performance of indefinite inner product refered to as IndSim($\mathbb{S}^n$), where we replace inner product in \eqref{eq:inner_prod} through symmetric positive semi-definite matrix $A^T A$ by an inner product through a nonsingular matrix $A$, thereby allowing similarity to take negative values. At last, we let CosSim($\mathbb{S}^n$) refer to the model where after the unit sphere projection we apply dot product. Note that since dot product is a special case of inner product, all these variants of defining similarity measure between two vectors can be succinctly described as optionally doing the unit sphere projection followed by a particular type of inner product.

\begin{figure}[!htbp]
\centering
\includegraphics[width=7.4cm]{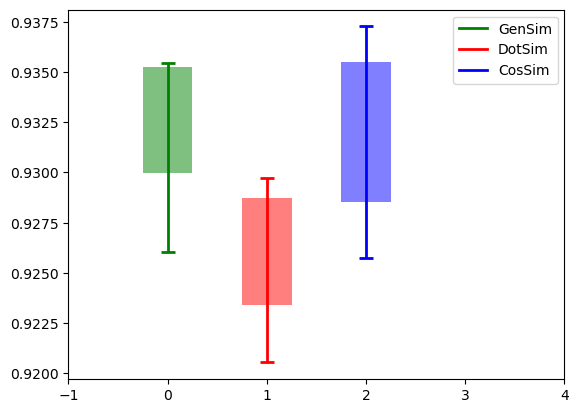}
\caption{Performance (AUC) and its $\mu$, $\sigma$ and $\rho$ for different similarity measures} 
\label{fig:auc_stat}
\end{figure}

\addtolength{\textheight}{-0.5cm} 

\begin{figure}[!htbp]
\centering
\begin{subfigure}{0.5\textwidth}
  \centering
  \includegraphics[width=6cm]{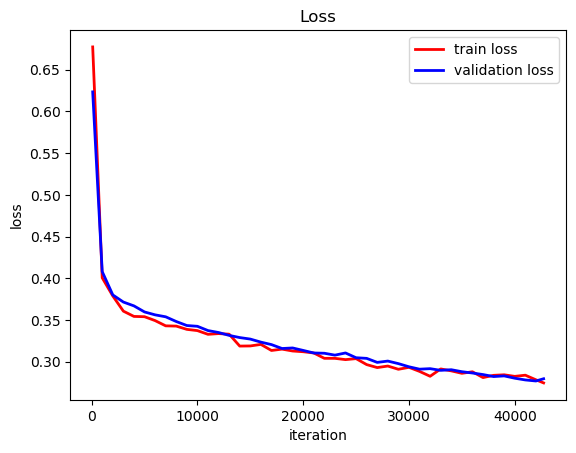}
  \caption{Train and Validation losses}
  \label{fig:loss}
\end{subfigure}%
\begin{subfigure}{0.5\textwidth}
  \centering
  \includegraphics[width=6cm]{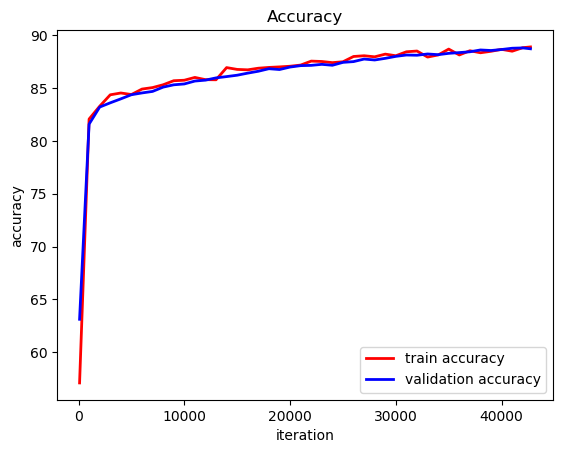}
  \caption{Train and Validation accuracies}
  \label{fig:acc}
\end{subfigure}
\caption{Train and validation loss and accuracy for GenSim($\mathbb{S}^n$) model} 
\label{fig:loss and acc}
\end{figure}

In order to compare GenSim($\mathbb{S}^n$), DotSim($\mathbb{R}^n$) and CosSim($\mathbb{S}^n$) similarity measures within TBSM, we record the corresponding mean $\mu$, standard deviations $\sigma$ and range $\rho$ of their AUC scores, which are $0.0023$ for the former and $0.0027$ for latter model. Assuming that AUC is normally distributed with given means and standard deviations the estimated chance that GenSim($\mathbb{S}^n$) performs better than DotSim($\mathbb{R}^n$) is $95\%$, as shown on Fig. \ref{fig:auc_stat}. Finally, the loss and accuracy obtained during training of the best GenSim($\mathbb{S}^n$) model are shown on Fig. \ref{fig:loss and acc}. Notice that the convergence shows the standard training L-shape curve, as expected.

Then, we experiment with different number and setup of TSLs. We let TSL (k-inner) designate a model with $k$ TSLs, each with its own inner product similarity measure on the unit sphere (after projection). We point out that GenSim($\mathbb{S}^n$) is a special case of TSL(k-inner) model with $k=1$. Moreover, we also test the use of different lengths subsequences. The idea of TSL(k-seq) approach is to directly amplify influence of most current items in the datapoint's time series. To achieve this effect we compute context vector not only for the full $\tau=20$ series but also for the most recent $\tau=5$, $\tau=10$ and $\tau=15$ subsequences. In this case, the output of TSL consists of four pairs $(\textbf{z}_{\tau},\textbf{c}_t)$, where $\textbf{c}_t$ is a context vector for the corresponding subsequence. This method can also be viewed as an example of ensemble technique, where instead of having different models we provide the same model with different data.

\begin{table}[!htbp]
\caption{Performance (AUC) on Taobao dataset}
\centering
\begin{tabular}{ | c | c | c| }
\hline
\hline
Model & Time Series Processing & AUC (average)  \\
\hline
TBSM & GenSim($\mathbb{S}^n$) & \setrow{\bfseries}0.9319    \\
TBSM & CosSim($\mathbb{S}^n$) & 0.9313    \\
TBSM & DotSim($\mathbb{R}^n$) & 0.9261  \\
TBSM & IndSim($\mathbb{S}^n$) & 0.9246 \\
\hline
TBSM & TSL(4 - seq) & 0.9279  \\
TBSM & TSL(8 - seq) & 0.9218  \\
TBSM & TSL(4-inner) & 0.9273  \\
TBSM & TSL(8-inner) & \setrow{\bfseries}  0.9327  \\
\hline
TBSM & LSTM(5-stack) & 0.8404  \\
TBSM & MHA (8-heads) & 0.8833   \\
\hline
\hline
MIMN & default & \setrow{\bfseries} 0.9179   \\
DIEN & default & 0.9081   \\
\hline
\end{tabular}
\label{auc}
\end{table}

Third, we compare the performance of TBSM with TSL we have proposed versus two standard mechanisms for processing time series data. The first mechanism uses standard multi-head attention (MHA) with 8 heads to replace the TSL. Note that the output of TSL is one pair $(\textbf{z}_{\tau},\textbf{c})$, where $\textbf{c}$ is the usual output of 8-head attention mechanism. Unlike the model with GenSim($\mathbb{S}^n$)  here we map time series vectors into $\mathbb{R}^n$ (as opposed to $\mathbb{S}^n$) using Key and Query matrices, and weight normalization is done at the end using softmax function ensuring that the sum of weights across time series is $1$. Note that this normalization is across time series, while in our main model normalization (projection onto $\mathbb{S}^n$) is done for each time series vector individually.

The last mechanism replaces TSL with recurrent neural network based on LSTM cells \cite{Hochreiter1997}. These cells are sized to input and output $n$ dimensional vectors and are stacked into 5 vertical layers. In this case the output is a pair $(\textbf{z}_{\tau},\textbf{c})$, where $\textbf{c}$ is the final hidden state. We remark that as compared to the default approach, LSTM model is not invariant under permutations of time series points.  

The performance achieved by all of these variations is summarized in the Tab. \ref{auc}. Notice that it is split into four parts following our experimental setup. The first part contains a class of smaller models, consisting of just one TSL. The best result in this class is achieved by the GenSim($\mathbb{S}^n$). The second class consists of larger models, the ones with more than one TSL. The overall best performing model TSL(8-inner) is in this class. The third class contains MHA- and LSTM-based models. Finally, we also add two well-known reference models MIMN \cite{qipi2019} and DIEN \cite{zhou2018deepi} for comparison. Note that in our experiments TBSM significantly outperforms both of them. 

\addtolength{\textheight}{-0.5cm} 

We remark that although performance of all models was calculated over the same test dataset, train dataset for models presented in this paper is different from the one used for MIMN and DIEN. Our training data consists of datapoints which are random contiguous subsequences of length $\tau=20$ extracted from each user behaviour history, while the dataset which was used to train MIMN and DIEN models consisted of datapoints of full length history sequences for each user $\tau=200$. Therefore, we may conclude that in our experiments TBSM achieves higher statistical performance than existing more complex models, all while using shorter sequence lengths.

%% file: related.tex
\section{Related Work}

Notice that click-through rate (CTR) prediction can be seen as a special case of matrix completion problem, where we predict an event probability and rating $r_{ui} \in \{0,1\}$ that represents clicks and non-clicks. The data corresponding to CTR problem may be broadly classified into two types based on whether each sample contains only user and item features or a time series corresponding to history of user activity. In this section we summarize modern approaches to solving the CTR prediction problem for both scenarios. 

The recommendation systems field received significant attention after the development of matrix factorization \cite{koren2009matrix}. The renewed focus on deep learning suggested the use of MLPs, rather than dot-products, as a function giving the final rating between user and item embeddings \cite{he2017neural}. Then, factorization machine method was developed in \cite{rendle2010} and further explored in \cite{guo2017deepfm, naumov2019deep, rendle2010}, while the autoencoder approach in \cite{kuchaiev2017training} provided another example of using neural networks for continuous rating prediction problem. Higher degree polynomial approximations of feature interactions were explored in  \cite{wang2017deep}, while self-attention was adopted in \cite{Song_2019}. In \cite{yang2019operationaware} different sparse feature embeddings are built depending on the consequent operations performed on these embedding, resulting in more than one embedding per feature. Lastly, \cite{Liu_2019} utilized convolutional neural networks combined with MLPs to learn new useful features. 

Along these developments the time component was introduced explicitly into some of these models. In \cite{wu2017} the time-dependent user-movie rating is decomposed into a sum of static and dynamic components, where first one is learned via factorization algorithm while the dynamic component is provided by LSTM \cite{Hochreiter1997}. The authors in \cite{beutel2017} incorporate contextual information such as time between successive user-video interactions into their RNN model. In \cite{chen2018wsdm} Memory-Augmented Neural Networks are used to learn the dynamics of factorization vectors used in matrix factorization. On the other hand, DIN \cite{zhou2018deep} enhanced basic embedding and MLPs with soft attention mechanism (interest) between the current item and user history. In DSIN \cite{feng2019deep} this idea is further enhanced by splitting user history into local sessions, while in DIEN \cite{zhou2018deepi} temporal dynamics are added to the concept of interest by using a variant of RNN. Finally, the most recent MIMN model \cite{qipi2019} learns from long history sequences using Neural Turing Machines for managing the memory network \cite{graves2014neural}. The TBSM model proposed in this paper incorporates several of these developments, but remains relatively simple, while retaining high statistical performance.

%% file: conclusion.tex
\section{Conclusion}\label{conclusion}

We have proposed time-based sequence model (TBSM) that incorporates embedding layer enhanced with TSL attention-like mechanism for handling time series data. In contrast to standard approaches, our approach to attention was geometric in a sense that attention weights came from similarity measure closely related to spherical distance function. 

We have shown on a synthetic dataset that as relationship between time series components gets stronger, adding different inner products helps model achieve a higher statistical score. Moreover, we have shown that taking relatively short time period ($\tau=20$) is sufficient for achieving a good statistical performance on Taobao User Behavior dataset, as measured by AUC metric. In our experiments TBSM has outperformed the scores attained by existing more complex models, all while using shorter sequence length.

Finally we point out that there is significant opportunity for exploring parallelism in TBSM, because embedding layer processes each location in time series separately and multiple TSLs are also independent of each other. We leave this exploration as future work.

\addtolength{\textheight}{+1.0cm} 